\documentclass[aps,prl,preprint,showpacs]{revtex4}
\usepackage{graphics,epsfig,amsmath,amssymb}
\begin{document}

\title{Entanglement in the Dicke model}
\author{S.~Schneider$^{1,2}$ and G.J. Milburn$^2$}
\affiliation{$^1$Department of Chemistry, University of Toronto, 80 St.~George
Street, Toronto, ON, M5S 3H6, Canada \\
$^2$Centre for Quantum
Computer Technology,
The University of Queensland, St.~Lucia, QLD 4072, Australia}

\begin{abstract}
We show how an ion trap, configured for the coherent manipulation of
external and internal quantum states, can be used to simulate the irreversible
dynamics of a collective angular momentum model known as the Dicke model.
In the special case of
two ions, we show that entanglement is created
in the coherently driven steady state with linear driving. For the case of
more than two ions we calculate the entanglement between two ions in the
steady state of the Dicke model by tracing over all the other ions. 
The entanglement in the steady state is a maximum for the
parameter values corresponding roughly to a bifurcation of a fixed point 
in the corresponding semiclassical dynamics.
We conjecture that this is a general mechanism for entanglement creation in
driven dissipative quantum systems.

\end{abstract}

\maketitle

\section{Introduction}

Generically, many body quantum systems
are known to be difficult to simulate efficiently on
a classical computer. This is because the quantum system may explore
regions of state space
with non zero entanglement, giving these systems access to a vastly larger
state space than is possible classically. In an open quantum system we may,
in some circumstances, be able to resort to stochastic methods, such as
Monte-Carlo
simulations. However, this will not be possible for open systems
in which the steady state itself is entangled, as in the example we
describe here.
Terhal {\em et al.}\cite{Terhal2000} have  considered the
possibility of using a quantum computer to simulate open quantum systems in
thermal equilibrium. Plenio {\em et al.}\cite{Plenio} have considered how decay
can lead to
entanglement rather than destroying it.
Cabrillo {\em et al.}\cite{Cabrillo} discuss creating entanglement
in two or more atoms, by driving  the atoms with a weak laser pulse
and detecting the spontaneous emission. In a recent paper by Arnesen {\em et
al.} \cite{Arnesen}, the authors look at a situation where the spins in a
Heisenberg chain with an external magnetic field show entanglement in the
thermal state with non-zero temperature. 
In this paper we consider the case
of an ion trap quantum computer realisation
to simulate the irreversible dynamics of $N$ two level systems.

In the following we will investigate the use of an ion trap quantum computer
as a simulation device for the so-called Dicke model \cite{Dicke:54}. This
model includes resonance fluorescence of a set of two-level atoms driven by a
resonant coherent laser field as well as a collective decay mechanism. We
will first describe
how a collective decay mechanism may be realised for $N$ trapped ions
interacting with
a collective vibrational mode when the vibrational mode is subject to
controlled heating.
Simulating irreversible dynamics for a trapped ion has been previously
suggested
by a number of authors \cite{Lutkenhaus}. 
We consider the dynamics of the
system relaxing from an initial state
in which all ions are in the excited state. For two ions we
explicitly calculate the Wooters entanglement measure
(concurrence)\cite{Wootters:98}
as a function of time, which shows that
the system passes through entangled states on the way to the collective
ground state. 
Next we add
coherent driving to try to force the system into a non trivial steady state
and show that, again for the
case of two ions, this steady state can be partially entangled. Extending
this result to many ions is
not possible at present due to the lack of a general measure of
entanglement of mixed states
in higher dimensions. However, we can calculate the entanglement between two
ions or atoms by tracing over all the other ions or atoms. This will show us
at least whether entanglement is present. 
Interestingly, the maximum entanglement
occurs for parameter values for which the corresponding semiclassical
system undergoes a bifurcation and loss of stability of the fixed point.
We conjecture that the loss of stability of a semiclassical fixed point
will generically be associated with
entanglement in the steady state of the full quantum system.

\section{The model}

In the seventies the Dicke model and cooperative effects were subjects of
research
in various groups (see e.g.~\cite{Walls:78,Drummond:78,Bowden} and references
therein).  The model consists of a group of two-level atoms
which is placed in a volume with dimensions small compared to
the wavelength associated with the atom's two-level dipole and evolves
on time scales
shorter than any $\hat{\mathbf{J}}^2$ breaking relaxation mechanism
(see \cite{Walls:78}) like an angular momentum system which has
collective atomic raising and lowering operators $\hat{J}_+$ and $\hat{J}_-$
with a fixed spin quantum number $j=N/2$, where $N$ is the number of atoms.

In the rotating frame with Markov, electric dipole and rotating wave
approximations and ignoring a small atomic frequency shift
the master equation for the density matrix
of this group of atoms under the cooperative influence of an electromagnetic
field is
\cite{Drummond:78,Drummond:80,Agarwal}
\begin{eqnarray}
\frac{\partial \hat{\rho}}{\partial t} & = &- i \frac{\Omega}{2}
\left[ \hat{J}_+
+ \hat{J}_-, \hat{\rho}\right]  \nonumber \\
&  & +{}  \frac{\gamma_{A}}{2} \left(2
\hat{J}_- \hat{\rho}\hat{J}_+ - \hat{J}_+ \hat{J}_- \hat{\rho} -
\hat{\rho}\hat{J}_+ \hat{J}_- \right) \,\, ,
\label{Dickemasterequation}
\end{eqnarray}
where $\Omega$ is the Rabi frequency and $\gamma_{A}$ is the Einstein
A-coefficient of each atom.
This model can be solved exactly \cite{Drummond:80} and it exhibits a
critical-point nonequilibrium phase transition for $\Omega/j = \gamma_{A}$ in
the limit $\Omega, j \to \infty$ \cite{Drummond:78}.

\subsection{Collective driving}

How do we get a similar master equation to Eq.~(\ref{Dickemasterequation}) in
an ion trap? The coherent evolution is easy: We just shine the same laser at
the carrier frequency on all the ions at the same time, thus forcing each ion
to undergo Rabi oscillations at the same frequency. If we start initially
with all the ions in their electronic ground state $|g\rangle$, the ions will
not leave the $j=N/2$ space. From there we can then define collective angular
momentum operators in the following way
\begin{eqnarray}
\hat{J}_- & = & \sum_{i=1}^N \hat{\sigma}_-^{(i)} \label{jminus}\\
\hat{J}_+ & = & \sum_{i=1}^N \hat{\sigma}_+^{(i)} \label{jplus} \,\, ,
\end{eqnarray}
where the raising and lowering operators for each ion are defined by
$\hat{\sigma}_- = |g \rangle \langle e|$ and $\hat{\sigma}_+ = |e \rangle
\langle g|$. With this the Hamiltonian for simultaneous resonant driving of
all the ions can be written as
\begin{equation}
\hat{H} = \hbar \frac{\Omega}{2} \left( \hat{J}_+ + \hat{J}_- \right) \,\, ,
\end{equation}
where $\Omega$ is the Rabi frequency for the electronic transition.

\subsection{Cooperative damping}

For the collective decay mechanism we
need to couple the ions equally to the same heat reservoir.
In this paper we will argue that the reservoir may be taken to be the
centre-of-mass vibrational mode. It is subject to heating and we
assume that it is in a thermal state. To couple the ions
to the vibrational mode we need another
laser which, again, illuminates all the ions at the same time, but which is
detuned from the carrier frequency to the red by the trap
frequency so that the electronic state of each atom gets coupled
simultaneously to the centre-of-mass vibrational mode.
This is described by a Hamiltonian for the $i$th ion
of the form
\begin{eqnarray}
\hat{H}_{\mathrm{red}}^{(i)} & = & \hbar \Omega_2  \left(
\hat{\sigma}_+^{(i)} \hat{a}
+ \hat{\sigma}_-^{(i)} \hat{a}^\dagger \right) \,\, ,
\end{eqnarray}
where we have introduced the bosonic annihilation operator $a$ for
the vibrational mode  and the coupling constant is
$ \Omega_2=\eta\Omega_0 $.
The parameter $\eta^2=E_r/(\hbar M\omega_0)$ is the Lambe-Dicke parameter
with $E_r$ the recoil kinetic energy of the atom, $\omega_0$
is the trap vibrational frequency, and $M$ is the effective mass for the
centre-of-mass mode. The Lamb-Dicke limit assumes
$\eta \ll  1$, which is easily achieved in practice. The frequency
$\Omega_0$ is the effective Rabi frequency for the
electronic transition involved.
This sideband transition is used to
efficiently remove
thermal energy from the vibrational degree of freedom. If the rate of this
cooling process can overcome
heating due to external fluctuations in the trap potential,
the ion may eventually be prepared in the vibrational ground state. However, 
in general the vibrational
state will reach a thermal mixture, $\hat{\rho}_v={\cal
Z}^{-1}\exp[-\hbar\omega_0 \hat{a}^\dagger \hat{a}/k_BT]$, where ${\cal Z} = 
\mathrm{Tr}(\exp[-\hbar\omega_0 \hat{a}^\dagger \hat{a}/k_BT])$ at some 
effective temperature $T$.
If the heating and cooling rates are such that the system relaxes at a
rate, $\alpha$, large
compared with any other time scale for ion motion, the ion can effectively
be regarded
as interacting with a thermal reservoir at temperature $T$.  We can also
arrange that the associated rate
of energy dissipation is small, $\gamma_A \ll\omega_0$  which simply requires
that the coupling
to the vibrational degree of freedom is weak. Finally we assume that the
temperature of the
vibrational degree of freedom is such that $\gamma_A \ll k_BT/\hbar$. Under
these assumptions
we may eliminate the description of the vibrational motion from the
dynamics and obtain a
master equation for the electronic state $\hat{\rho}^{(i)}i$ of the ion
\begin{eqnarray}
\frac{d\hat{\rho}^{(i)}}{dt} & = &-\frac{i}{\hbar}[\hat{H},\hat{\rho}^{(i)}]
      +\gamma_A\bar{n}{\cal D}[\hat{\sigma}_+^{(i)}]\hat{\rho}^{(i)}
     \mbox{}+\gamma_A(\bar{n}+1){\cal
D}[\hat{\sigma}_-^{(i)}]\hat{\rho}^{(i)}\, \, ,
\label{single_me}
\end{eqnarray}
where the superoperator is defined by
\begin{equation}
{\cal D}[\hat{A}]\hat{\rho}=\hat{A}\rho\hat{A}^\dagger
-\frac{1}{2}(\hat{A}^\dagger \hat{A}\hat{\rho}+\hat{\rho}\hat{A}^\dagger
\hat{A})\ ,
\end{equation}
and where $\hat{H}$ is the Hamiltonian for any other reversible electronic
dynamics
and $\bar{n}$ is the mean thermal occupation number of the
vibrational degree of freedom. In what follows we assume that the cooling is
very efficient and set $\bar{n}=0$.
At any time we my turn off the cooling lasers, thus reducing $\gamma_A$
suddenly to zero. The master equation, Eq.(\ref{single_me}),
looks like the master equation for a single two level atom interacting with
the many body radiation field via spontaneous emission and thermal
absorption and we will discuss this case in the next section. In the case
considered here, however, we ignore
spontaneous emission which is justified if the electronic states coupled by
the lasers correspond to a long lived
dipole forbidden transition. Thus the irreversible dynamics of the
electronic state is due entirely to
the interaction with the phonons associated with the vibrational degree of
freedom.

If the external laser field on each ion is identical (in amplitude and phase)
the interaction Hamiltonian is
\begin{equation}
\hat{H}_I=\hbar\Omega_2(\hat{a}\hat{J}_+ +\hat{a}^\dagger \hat{J}_-) \,\, , 
\label{TC}
\end{equation}
where $\hat{J}_+$ and $\hat{J}_-$ are defined in Eqs.~(\ref{jplus}) and
(\ref{jminus}).
For the case of a linear ion trap, with separately
addressable ions, identical laser
fields could easily be obtained by splitting the cooling laser into
multiple beams.
In this way we can simulate an angular momentum system with quantum number
$j=N/2$. This imposes a permutation symmetry on the system which reduces
the effective
Hilbert space dimension from $2^N$ to $2N+1$. Thus an
exponentially large portion of the available Hilbert space, i.e.~all the
states with $j<N/2$,
is not used in this simulation.
However, it is easy to generate the relevant unitary transformations
to simulate the $j=N/2$ angular momentum quantum system.

It is not trivial to keep the vibrational mode in a thermal state of fixed
temperature. One way of doing this was recently suggested by Kielpinski {\em
et al.}~\cite{Kielpinski:99}: They put one ion, which is a different species
of all the other ions, in the centre of a string of ions, so that they have an
odd number of ions in the trap. Through this centre ion, which they can cool
at will without disturbing the coherence of the other ions, the
sympathetically cool all the ions and are thus able to keep the string of ions
at a well defined temperature. The authors conclude that such a scheme of
sympathetic cooling is ``well within the reach of current experimental
technique'' \cite{Kielpinski:99}. We assume for further calculations that the
centre-of-mass mode is kept in such a thermal state by the outlined technique.
In the following derivation of the master equation we do not explicitly put
the laser cooling into the equation. We just assume that the vibrational state
instantly, i.e.~on a time scale fast compared to all the other processes
involved, relaxes back into the thermal state.

With all these assumptions we get the master equation
describing the collective motion of the density
matrix of all the ions
\begin{eqnarray}
\frac{\partial \hat{\rho}}{\partial t} & = & = - i \frac{\Omega}{2}
\left[ \hat{J}_+ + \hat{J}_-, \hat{\rho}\right] \nonumber \\
&  & + {} \gamma_{A} \frac{\bar{n}}{2}
\left(2 \hat{J}_+ \hat{\rho} \hat{J}_- - \hat{J}_- \hat{J}_+ \hat{\rho} -
\hat{\rho} \hat{J}_- \hat{J}_+ \right)\nonumber \\
&  &  {}+ {}\gamma_{A} \frac{\bar{n}+ 1}{2}
\left(2 \hat{J}_- \hat{\rho}\hat{J}_+ - \hat{J}_+ \hat{J}_- \hat{\rho} -
\hat{\rho}\hat{J}_+ \hat{J}_- \right)  \,\, ,
\label{mastereqlargeangdensity}
\end{eqnarray}
where $\bar{n}$ is the mean phonon number of the vibrational centre-of-mass
mode and $\gamma_{A} = 2 \Omega_2^2 \eta^2 $. Note that for $\bar{n} = 0$ this
equation is identical to Eq.~(\ref{Dickemasterequation}) for the Dicke model.

\section{Steady state and entanglement for $j=1$}

With two ions we have
$j=1$ and from the master equation, Eq.~(\ref{mastereqlargeangdensity}), we
can write down the equation of motion for the components of the
three by three density matrix of the state of the system, taking into
account that $\mathrm{Tr}(\hat{\rho}) = 1$ and that $\hat{\rho}$ is Hermitian.
Getting the steady state is then a matter of simple algebra.

Once we have
determined the steady state of the $j=1$ system, we can rewrite this state in
the underlying two-qubit basis. What we are interested in is the change of
entanglement in the system as the parameters $\gamma$ and $\bar{n}$ change.
The entanglement of two qubits is well defined
\cite{Bennett:96,Wootters:98,Vedral:98},
and we choose the concurrence \cite{Wootters:98} as a measure
for it.

\begin{figure}[hhh]
\centering
\resizebox{7cm}{!}{\includegraphics{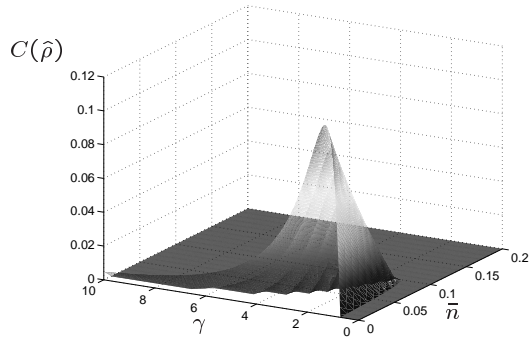}}
\caption{Plot of the concurrence as a measure of
entanglement  depending on the
parameters $\gamma = \gamma_{A}/\Omega$ and $\bar{n}$, the mean phonon number
of the thermal vibrational state. }
\label{concurrenceplot}
\end{figure}

A numeric evaluation of the concurrence leads to the
plot in Fig.~\ref{concurrenceplot}.
What we see is that we can get a certain amount of entanglement in the steady
state of a coherently driven system that is coupled to a thermal reservoir.
This is remarkable as the steady state is independent of the initial state,
which can be unentangled. The coherent evolution alone does not lead to any
entanglement for an initially unentangled state either, as it only consists of
(simultaneously acting) single qubit rotations and no coupling between the
qubits is present. Thus the entanglement
is due to the cooperative decoherence in the system acting together with the
coherent evolution.

\section{Zero temperature case.}

The analysis here is restricted to the case $j=1$. For $j>1$ and
$\bar{n}\neq 0$, numerical methods
will need to be employed to derive the steady state, however,
in this case another problem will
arise due to the fact that there is currently no measure of entanglement for
$N$ coupled qubits. Nevertheless, other phase transitions analogous to the one
in the Dicke-model (see e.g.~\cite{Walls:78,Drummond:78} and references
therein) will appear.

For $\bar{n} =0$ we can compare our results for the steady state to the ones
calculated by Puri and Lawande \cite{Puri:79} (see also Lawande
{\em et al.}~\cite{Lawande:81}).
They calculate the steady state to be
\begin{equation}
\hat{\rho}_{S} = \frac{1}{D} \sum_{m,n =0}^2 (g^\ast)^{-m} (g)^{-n}
\hat{J}_-^m \hat{J}_+^n \,\, ,
\end{equation}
where
\begin{equation}
D= \sum_{k=0}^2 H_{2,k} |g|^{-2k}
\end{equation}
is a normalization constant, $g = i /\gamma$, where $\gamma =
\gamma_{A}/\Omega$ as defined above, and
\begin{equation}
H_{2,m} = \frac{(2 + m + 1)! (m!)^2}{(2-m)! (2m+1)!} \,\, .
\end{equation}
With this we can write the density matrix of the steady state in matrix form
as
\begin{equation}
\hat{\rho}_{S} = \frac{1}{D} 
\left( \begin{array}{ccc} 1 & - i\sqrt{2}  \gamma & - 2 \gamma^2 \\
i \sqrt{2} \gamma & 1 + 2 \gamma^2 & -i \sqrt{2} \gamma -i 2 \sqrt{2} \gamma^3
\\
-2 \gamma^2 & i \sqrt{2} \gamma + i 2 \sqrt{2} \gamma^3 & 1 + 2 \gamma^2 + 4
\gamma^4
\end{array} \right) \,\, ,
\end{equation}
where we have calculated $D$ as
\begin{equation}
D = 3 + 4 \gamma^2 + 4 \gamma^4 \,\, .
\end{equation}
\begin{figure}[hhh]
\centering
\resizebox{7cm}{!}{\includegraphics{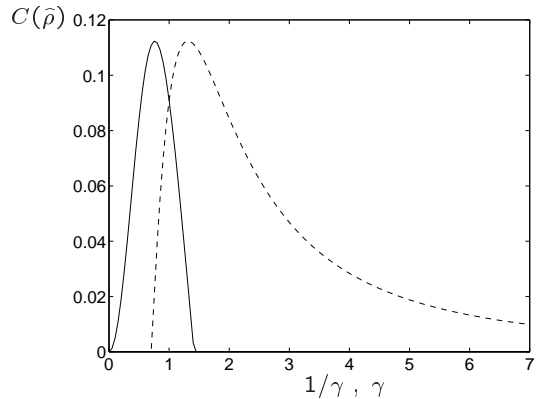}}
\caption{Plot of the concurrence as a measure of
entanglement  depending on the
parameters $|g|= 1/ \gamma = \Omega/\gamma_{A}$ and $1/|g| = \gamma$ (dashed
line) with $\bar{n}=0$, i.e.~for the Dicke-model.}
\label{nzerocon}
\end{figure}
This gives the same result as above if we set $\bar{n} =0$.
The concurrence for this special case is plotted in figure \ref{nzerocon},
but this time we plot it against $|g| = 1/\gamma$ as well. Thus the
Dicke-model as a special case of our model shows entanglement in the steady
state.

\begin{figure}[hhh]
\centering
\resizebox{7cm}{!}{\includegraphics{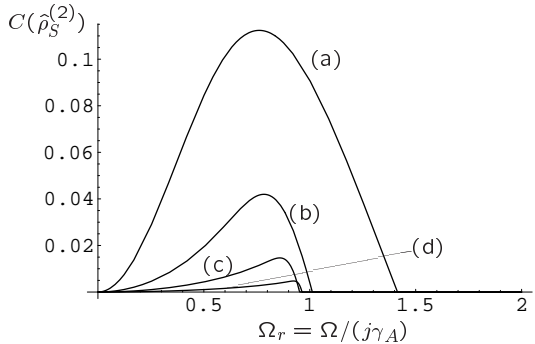}}
\caption{Plot of the two-ion concurrence as a measure of
entanglement depending on the
parameters $\Omega_r= \Omega/(j \gamma_{A})$ for (a) $j=1$, (b) $j=4$, (c)
$j=16$, and (d) $j=64$.}
\label{concplot}
\end{figure}

For $N>2$ up to date no definite measure of entanglement exists. But we can
calculate the entanglement between just two of the $N$ ions at least for the
Dicke model, where the temperature of the bath is zero. The steady is then
given by \cite{Puri:79, Lawande:81, Drummond:80}
\begin{equation}
\hat{\rho}_S = \frac{1}{D} \sum_{l=0}^{2 j} \sum_{l^\prime = 0}^{2 j}
\left(\frac{\hat{J}_-}{g} \right)^l \left(
\frac{\hat{J}_+}{g^\ast}\right)^{l^\prime} \,\, . 
\end{equation}
By writing this as a sum of states with angular momentum $j_1 = 1 $ and $j_2 =
j-1$ we can trace over the part of the Hilbert space with $j_2 = j-1$ and thus
get the density matrix in the steady state for just two ions (or atoms in the
original Dicke model). From there we can again calculate the concurrene. This
time we plot it against the relative Rabi frequency \cite{Drummond:80}
$\Omega_r = \Omega/(j \gamma)$. We note that the maxima of the entanglement
occur close to the critical point in the cooperative limit $j, \Omega
\rightarrow \infty$ of the Dicke model, i.e.~around $\Omega_r = 1$. The
two-ion entanglement is not the real measure of entanglement in the system.
Thus we cannot take the cooperative limit as the two-ion entanglement goes to
zero in this limit. However, we note from the plots in Fig.~(\ref{concplot})
that the maximum value of the two-ion entanglement indeed does move closer to
the point $\Omega_r = 1$ for increasing $N$. 

\section{Conclusion}
In this paper we have demonstrated how the steady state of a dissipative
many body system,
driven far from equilibrium, may exhibit non zero quantum entanglement.
This result is significant for two reasons.
Firstly the steady state is a mixed state and the study of quantum
entanglement for mixed states is a very active field of
enquiry \cite{Kraus2000}. It
immediately raises the question of whether the entanglement can be distilled
and used as a resource for some quantum communication or computation
task \cite{Nielsen2000}.
Secondly the maximum entanglement occurs at the same parameter values for
which
the semiclassical dynamics of the system undergoes a bifurcation of the
fixed point
corresponding to the quantum steady state. At the bifurcation point the
time constant
associated with the fixed point goes to zero as the bifurcation is
approached. This is
reminiscent of a phenomenon that characterises quantum phase transitions,
in which a morphological
change in the ground state, as a parameter is varied, is associated with a
frequency gap tending to zero \cite{Sachdev2000,Milburn2000}.
We conjecture that the association between
the bifurcation of a fixed point of the semiclassical description and the
maximum of entanglement
will be a general feature of dissipative many body systems driven far from
equilibrium.

\section{Acknowledgments} The authors would like to thank Daniel James for
interesting discussions.

\end{document}